\documentclass[11pt,a4paper, final, twoside]{article}
\usepackage{amsmath}
\usepackage{fancyhdr}
\usepackage{amsthm}
\usepackage{amsfonts}
\usepackage{amssymb}
\usepackage{amscd}
\usepackage{latexsym}
\usepackage{amsthm}
\usepackage{graphicx}
\usepackage{graphics}
\usepackage{natbib}
\usepackage{afterpage}
\usepackage[colorlinks=true, urlcolor=blue,  linkcolor=black, citecolor=black]{hyperref}
\usepackage{color}
\numberwithin{equation}{section}
\fancypagestyle{plain}{
\fancyhf{} 

\fancyhead[C]
{\bf {Asian Journal of Research and Reviews in Physics}
\\\hspace*{-1.2cm}
{\quad\indent\quad\indent\quad\indent\quad\indent\quad\indent\quad\indent Volume 8, Issue 2, Page 66-72, 2024; Article no.AJR2P.127097}\\
\hspace*{-2cm}\hspace*{-2cm}\hspace*{-3cm} {\footnotesize{ \it{~ISSN: 2582-5992}}}\\
}
}
\begin{document}
\hyphenpenalty=100000

\begin{flushright}
\thispagestyle{plain} 

{\Large \textbf{\\Global Thermal Instability in the Spherical Interstellar Clouds}}\\[5mm]
{\large \textbf{M. Nejad-Asghar\footnote{\emph{E-mail: nejadasghar@umz.ac.ir}} }}\\[1mm]
{\footnotesize{\it Department of Theoretical Physics,\\ University of Mazandaran,
\\Babolsar, Iran}}
\end{flushright}

\begin{flushleft}\fbox{%
\begin{minipage}{1.8in}
{\slshape \textbf{Original Research Article (preprint)}\/}
\end{minipage}}
\end{flushleft}

\begin{flushright}\footnotesize \it Received: 14/09/2024\\
Accepted: 17/11/2024\\
Online Ready: 25/11/2024
\end{flushright}

{\Large \textbf{Abstract}}\\[4mm]
\fbox{%
\begin{minipage}{5.4in}{\footnotesize{Thermal instability (TI) is a trigger mechanism, which can explain the formation of small condensations through some regions of the interstellar clouds. The instability criterion for flat geometry approximations has been investigated in previous works. Here, we focus on spherical perturbations in the spherical clouds. Our goal here is to examine the conditions for the occurrence of TI through the thermally dominated (i.e., gravitationally stable) quasi-static spherical interstellar clouds. First, we obtain the profiles of density, temperature, pressure, and enclosed mass of a symmetric spherical cloud. Then, we use the perturbation method to investigate the linear regime of instability and find its growth rate.
Considering spherical perturbations on the quasi-static spherical cloud, instead of a thermal and dynamical equilibrium flat cloud, changes the instability criterion so that we can conclude that sphericalness can increase the occurrence of TI. The results show that in the spherical clouds, perturbations with shorter wavelengths have more chance to grow via TI (i.e., greater growth rates).}
} \end{minipage}}\\[1mm]

\small

\section{Introduction}\label{I1}
After the pioneer paper of [1], entitled \textit{thermal instability} (TI), this subject appeared to be an important mechanism to explain the formation of density condensations through interstellar and intergalactic clouds. This mechanism is used to explain a wide range of multiphase phenomena from local intra-galactic situations to large extra-galactic occurrences. For example, the stability of a uniform star-forming medium is discussed by [2] including thermal and magnetic effects. [3] revisited the effect of the ion-neutral friction of the two fluids on the growth of TI. The effect of the application of isobaric TI in forming the low-mass condensations within molecular cloud cores is investigated by [4]. Global linear stability analysis and idealized numerical simulations in global thermal balance are performed by [5] to understand the condensation of cold gas from hot/Virial atmospheres, in particular the intracluster medium. [6] investigated some conditions for the occurrence of TI and the formation of pre-condensations through the outer half of a quasi-static spherical molecular clump or core. The clumpy wind simulations in plasma of thermally unstable active galactic nuclei are presented by [7], obtained by simulating parsec-scale outflows irradiated and heated by X-rays. [8] deduced that during the cooling and driven by TI, solar coronal rain is produced along the loops. [9] aimed to show the role of TI as a constraint for warm, optically thick X-ray coronas in active galactic nuclei. Recently, [10] determined the effect of background flow on TI in cylindrical magnetic field configurations.

Due to the importance of TI as a cold gas formation mechanism, here, we investigate the examination for the occurrence of linear TI in the spherical interstellar clouds. The most important parameter for the occurrence of TI is the net cooling function $\Omega(\rho,T)$. In a local thermal and dynamical equilibrium flat clouds, the isobaric instability criterion is $\Omega_T - (\frac{\rho_0}{T_0})\Omega_\rho < 0$, where $\Omega_\rho \equiv (\partial \Omega / \partial \rho)_T$ and $\Omega_T \equiv (\partial \Omega / \partial T)_\rho$ are evaluated in equilibrium state. Using the flat plane perturbations to investigate local TI can be appropriate in some interstellar gases, but, many interstellar clouds have a spherical structure. Thus, the flat plane approximation is not completely correct for global TI, and it is better to use spherical perturbations.

For this purpose, in \S~2 we consider thermally equilibrium models for quasi-static spherically symmetric clouds. The effect of spherical perturbations is investigated in \S3. Section~4 is devoted to summary and conclusions.

\section{Thermally equilibrium models}
In the spherical polar coordinates, the usual hydrodynamic equations
for spherically symmetric thermally dominated clouds are
\begin{equation}\label{contin}
    \frac{\partial \rho}{\partial t} + \frac{1}{r^2} \frac{\partial}{\partial
    r} \left( r^2 \rho u \right) = 0,
\end{equation}
\begin{equation}\label{moment}
    \frac{\partial u}{\partial t} + u\frac{\partial u}{\partial r}
    =-\frac{1}{\rho} \frac{\partial p}{\partial r} - \frac{GM}{r^2},
\end{equation}
\begin{equation}\label{mass}
    \frac{\partial M}{\partial t} + u \frac{\partial M}{\partial r}
    = 0,\quad \frac{\partial M}{\partial r} = 4 \pi r^2 \rho,
\end{equation}
\begin{equation}\label{energy}
    \frac{\partial p}{\partial t} + u \frac{\partial p}{\partial r}
    + \gamma p \frac{1}{r^2} \frac{\partial}{\partial r} \left( r^2 u
    \right) = - \left( \gamma -1 \right) \rho \Omega,
\end{equation}
\begin{equation}\label{state}
    p = \frac{k_B}{\mu m_H} \rho T,
\end{equation}
where mass density $\rho$, the enclosed mass $M$, radial flow
velocity $u$, thermal gas pressure $p$ and temperature $T$ depend on
the radius $r$ and time $t$; $G$ is the gravitational constant,
$\gamma$ is the heat capacity ratio, and $k_B$, $\mu$ and $m_H$ are
Boltzmann constant, the mean particle weight and the hydrogen mass,
respectively. The net cooling function is presented by
$\Omega(\rho,T)= \rho^2 \Lambda(T)- \rho \Gamma$ where $\Lambda$ and
$\Gamma$ are the cooling and heating rates, respectively.

Determination of the net cooling function $\Omega$ for optically
thick and/or optically thin interstellar gas is a complex non-local
thermodynamic equilibrium radiative transfer problem. For example,
[4] used the results of [11] to parameterize the cooling rate for
molecular clouds as $\propto \left( T/10\,\mathrm{K} \right)^\beta$
where the parameter $\beta$ and proportional constant are given in
the figure~1 of his paper. For heating mechanisms, he considered
different heating mechanisms such as heating due to cosmic rays
(e.g., [12]), dissipation of magnetic energy via ambipolar diffusion
(e.g.,[13]), and so on. As another example, [14] used the data of
[15] to approximate an analytic function for the cooling rate in the
circumgalactic medium as
\begin{equation}\label{coolr}
\Lambda(T)= 3.9 \times 10^{26} \times
10^{\Theta(\log(T/10^5\,\mathrm{K})} \,\mathrm{erg}\,
\mathrm{cm}^3\, \mathrm{g}^{-2}\, \mathrm{s}^{-1},
\end{equation}
where the exponent is
\begin{equation}\label{coolexp}
\Theta(x) = 0.4 x -3 + \frac{5.2}{e^{x+0.08}+e^{-1.5(x+0.08)}}.
\end{equation}
For hating rate, they used a constant value $\Gamma= 0.06
\,\mathrm{erg}\, \mathrm{g}^{-1}\, \mathrm{s}^{-1}$, which is
expected from photoelectric explosion of electrons from dust grains.

In the stationary ($\partial/\partial t =0$) quasi-static
($u\rightarrow 0$) thermally equilibrium state, the net cooling
function $\Omega(\rho,T)$ is assumed to be zero at each radius $r$
(i.e., locally thermal balance). This thermal balance (i.e.,
$\Omega(\rho,T) = 0$) leads to a relation between the temperature
$T$ and density $\rho$ at each radius. Here, we use a parametric
relation between density and temperature as $T = \rho^\eta$, where
$\eta$ is a constant parameter. We consider three state-of-the-art
models for temperature change according to the density decrease of
the cloud: (i) decreasing temperature by $\eta>0$, (ii) increasing
temperature by $\eta>0$, and constant temperature by $\eta=0$.

\begin{figure}[h]
\begin{center}
\includegraphics[scale=0.7]{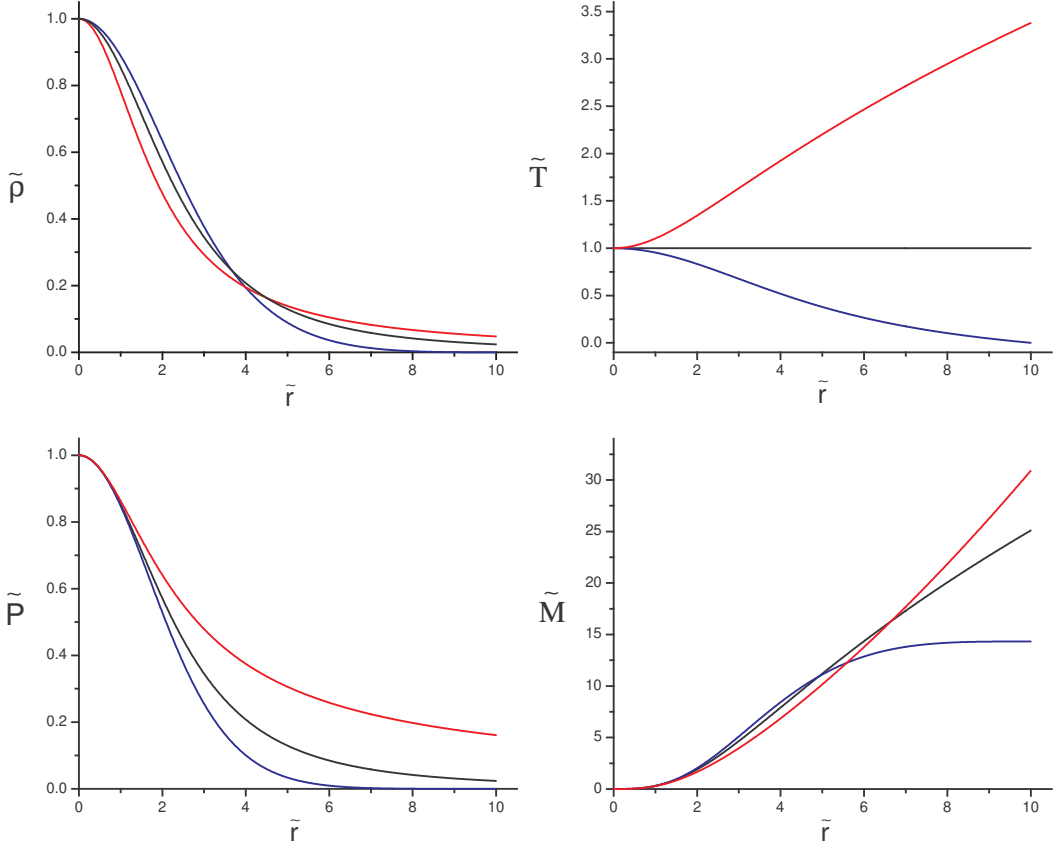}
\caption{\footnotesize{The thermally equilibrium profiles of
non-dimensional density, temperature, pressure, and enclosed mass
for a quasi-static spherically symmetric interstellar cloud. The
solid black curves are for $\eta=0$, blue curves are for $\eta=0.4$,
and red ones are for $\eta=-0.4$.}\label{prof}}
\end{center}
\end{figure}

Knowing the relation between temperature and density, equation
(\ref{state}) leads us to determine the gradient of pressure as
\begin{equation}\label{dpdr}
    \frac{dp}{dr} = \frac{k_B}{\mu m_H} \left( T+\rho\frac{dT}{d\rho} \right)
    \frac{d\rho}{dr},
\end{equation}
so that the stationary ($\partial/\partial t =0$) quasi-static
($u\rightarrow 0$) state of the momentum equation (\ref{moment})
becomes
\begin{equation}\label{hydrost}
    \frac{d\rho}{dr} = - \frac{\mu m_H G}{k_B}
    \frac{M\rho}{r^2 \left( T+ \rho\frac{dT}{d\rho} \right)}.
\end{equation}
We use the non-dimensional quantities $\tilde{\rho} \equiv
\rho/\rho_r$, $\tilde{T} \equiv T/T_r$, $\tilde{r} \equiv
r/(\frac{k_BT_r/\mu m_H}{4\pi G \rho_r})^{\frac{1}{2}}$, and
$\tilde{M} \equiv M/ 4\pi (\frac{k_BT_r/\mu m_H}{4\pi G
\rho_r})^{\frac{3}{2}} \rho_r$, where $\rho_r$ and $T_r$ are the
reference density and temperature, respectively. In this way, the
equations (\ref{mass}) and (\ref{hydrost}) become
\begin{equation}\label{hydmass}
    \frac{d\tilde{M}}{d \tilde{r}} = \tilde{r}^2 \tilde{\rho},
\end{equation}
\begin{equation}\label{hydmoment}
    \frac{d\tilde{\rho}}{d\tilde{r}} = - \frac{\tilde{M}\tilde{\rho}^{1-\eta}}
    {\tilde{r}^2 \tilde{T} (1+ \eta)},
\end{equation}
where the thermal balance relation $\tilde{T} = \tilde{\rho}^\eta$
is used. The differential equations (\ref{hydmass}) and
(\ref{hydmoment}) can be integrated numerically (e.g., with
Runge-Kutta method), from the origin $\tilde{r}=0$ with the boundary
conditions $\tilde{\rho}(0)=\tilde{\rho}_c$ and $\tilde{M}(0)=0$,
where $\tilde{\rho}_c$ is the non-dimensional central density. The
density profiles of some models with different values of the
parameter $\eta$ are depicted in Fig.~\ref{prof}.

\section{Perturbations in the cloud}
The quasi-static spherical interstellar clouds with low-density contrast are mainly confined by the external pressure. If the density contrast exceeds a critical value (e.g., Bonnor-Ebert sphere), the cloud will have gravity-dominated configurations, and an arbitrarily small initial perturbation in the structure grows rapidly with time, leading ultimately to collapse. Here, we turn our attention to the gravitationally stable cloud with masses less than the critical value of the Bonnor-Ebert sphere. If perturbations occur in the structure of these clouds, the thermal effects will be important for growth (i.e., thermal instability) or decay (i.e., thermal stability) of them. To investigate this effect, we split each variable into unperturbed and perturbed components; the latter is indicated by subscript '1'. We perform the linear perturbation analysis, with time and radius Fourier expansions, $A_1(r,t) = A_1^d \exp (\omega t + ikr)$, on the thermal equilibrium spherically symmetric cloud (in which its variables are denoted by subscript '0'). Time evolution in the non-linear regime is out of the scope of this paper. It is of great interest to derive the growth/decay rate, $\Re(\omega)$, for the range of suitable wavelengths, $2\pi/k$, at different radii $r$.

To investigate the thermal stability/instability of the medium, the linear perturbation method can be used. In this method, a perturbation is applied to the thermal equilibrium medium, and only the first order of domains is considered. If $\Re(\omega)$ is positive, the gas will be thermally unstable, and if $\Re(\omega)$ is negative, the perturbation will be decayed so that the gas will be thermally stable. In this way, by using perturbations of the form $\exp (\omega t + ikr)$ for density, radial velocity, enclosed mass, and pressure with amplitudes $\rho_1^d$, $u_1^d$, $M_1^d$, and $p_1^d$, respectively, the equations (\ref{contin})-(\ref{state}) can be linearized by repeated use of the unperturbed equations as follows
\begin{equation}\label{linear1}
    \left( \omega \right) \rho_1^d
    + \left( \rho_0' + \frac{2}{r}\rho_0 + ik \rho_0 \right) u_1^d = 0,
\end{equation}
\begin{equation}\label{linear2}
    \left( -\frac{p_0'}{\rho^2} \right) \rho_1^d + \left( \omega \right) u_1^d +\left( \frac{G}{r^2} \right) M_1^d + \left( \frac{ik}{\rho_0} \right) p_1^d = 0,
\end{equation}
\begin{equation}\label{linear3}
    \left(  4\pi r^2 \rho_0 \right) u_1^d +\left( \omega \right) M_1^d  = 0,
\end{equation}
\begin{equation}\label{linear4}
    \left[  \frac{p_0}{\rho_0} \left( \nu_\rho - \nu_T \right) \right] \rho_1^d
    + \left( p_0'  +\frac{2\gamma}{r} p_0 + ik\gamma p_0 \right) u_1^d
    + \left( \omega + \nu_T \right) p_1^d = 0,
\end{equation}
where $\nu_T\equiv \frac{\mu m_H (\gamma-1)}{k_B} \left(
\frac{\partial \Omega}{\partial T} \right)_\rho$, $\nu_\rho \equiv
\frac{\mu m_H (\gamma-1)}{k_B} \frac{\rho_0}{T_0} \left(
\frac{\partial \Omega}{\partial \rho} \right)_T$ are angular
frequencies of isochoric and isothermal perturbations, respectively
[1], and primes denote $\frac{d}{dr}$. Now, if we set the
determinant of coefficients matrix equal to zero, we obtain a third
degree polynomial characteristic equation as follows
\begin{equation}\label{charac}
    \omega^3 + C_2 \omega^2 + C_1 \omega + C_0 = 0,
\end{equation}
where
\begin{equation}\label{coeff0}
   C_0 = \left( - 4\pi G \rho_0 +\frac{p_0' \rho_0'}{\rho_0^2} +\frac{2}{r} \frac{p_0'}{\rho_0}
   + i k \frac{p_0'}{\rho_0} \right) \nu_T
   + \left( i k \frac{p_0 \rho_0'}{\rho_0^2} + \frac{2 i k}{r} \frac{p_0}{\rho_0} - k^2 \frac{p_0}{\rho_0} \right) (\nu_\rho - \nu_T),
\end{equation}
\begin{equation}\label{coeff1}
    C_1 = -4\pi G \rho_0- \frac{2 i k \gamma}{r} \frac{p_0}{\rho_0}
    +k^2 \gamma \frac{p_0}{\rho_0} + \frac{p_0' \rho_0'}{\rho_0^2} + \frac{2}{r} \frac{p_0'}{\rho_0},
\end{equation}
and $C_2 = \nu_T$. These coefficients depend on the wavenumber $k$,
and must be evaluated for different radii $r$.

\begin{figure}[h]
\begin{center}
\includegraphics[scale=0.7]{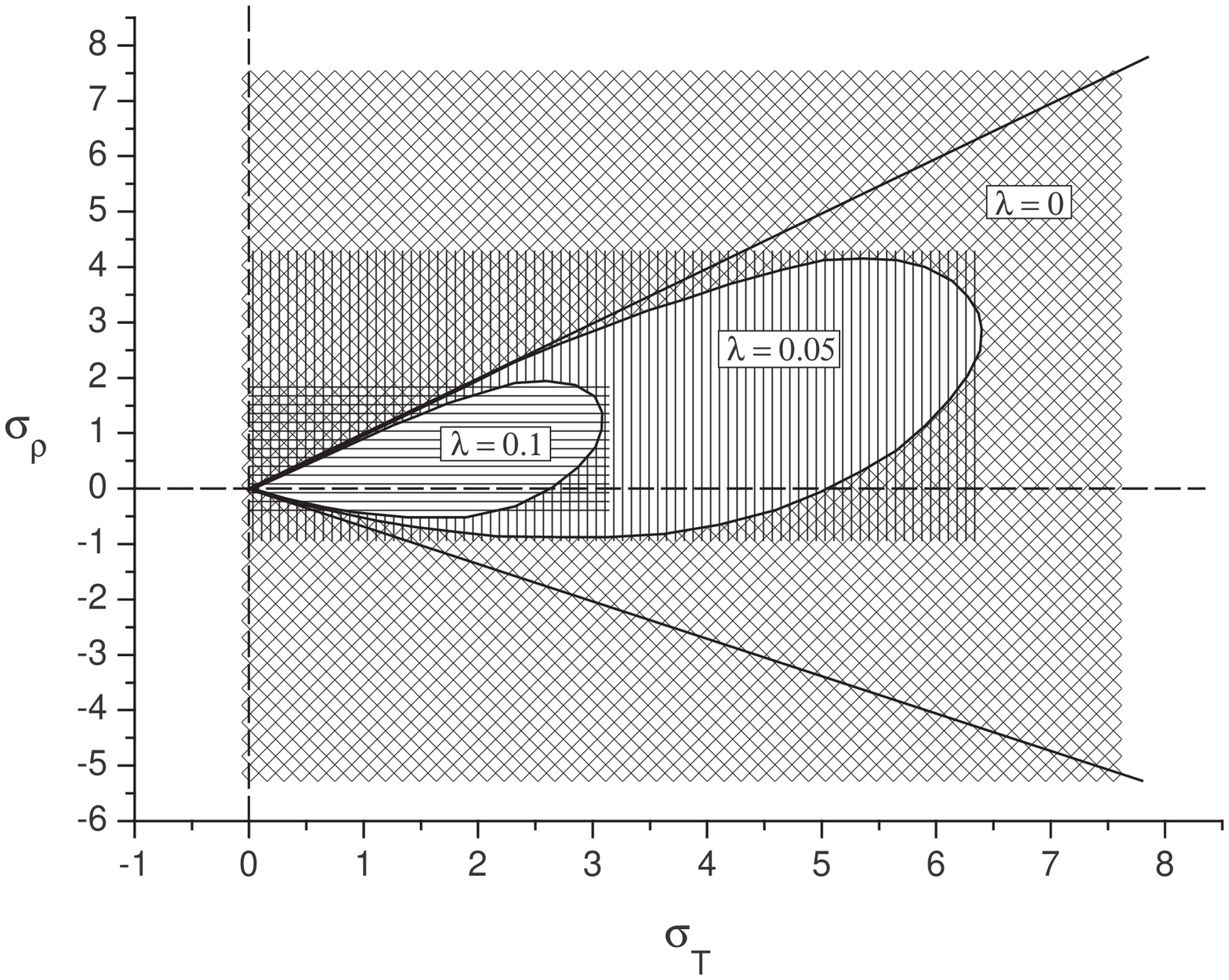}
\caption{\footnotesize{Regions of TI in the $\sigma_T$-$\sigma_\rho$
plane with $\lambda=0$, $0.05$ and $0.1$ for $\gamma=5/3$. The
stable regions are depicted by hatch-patters of diagonal cross for
$\lambda=0$, light vertical lines for $\lambda=0.05$, and light
horizontal lines for $\lambda=0.1$. In each case of $\lambda$, other
regions are thermally unstable.}\label{sisfig}}
\end{center}
\end{figure}

The simplest case is one in which $\eta=0$. Here, we turn our attention to this case, and other models, which depend strictly on the choosing of the net cooling function, will be considered in the subsequent research. In this case (i.e., $\eta=0$), the density and pressure profiles can be locally (i.e., regions of interest around each radius far from $r\approx0$) approximated as inverse square law, $p_0\, \& \rho_0 \propto \frac{1}{r^2}$, and the enclosed mass increases linearly as $M \propto r$. These results can be deduced from Fig.~\ref{prof}. In this way, the coefficients of the characteristic equation (\ref{charac}) reduce to
\begin{equation}\label{coeff00}
    C_0 = \frac{k^2 \bar{c}^2}{\gamma} (\nu_T - \nu_\rho) - i k \frac{2\bar{c}^2\nu_T}{\gamma r}
    - \frac{2\bar{c}^2\nu_T}{\gamma r^2},
\end{equation}
\begin{equation}\label{coeff11}
    C_1 = k^2 \bar{c}^2 - i k \frac{2\bar{c}^2}{r} - \frac{2\bar{c}^2}{\gamma r^2},
\end{equation}
and $C_2 = \nu_T$, where $\bar{c}$ is the sound speed. Note that the
terms including $1/r$ and $1/r^2$ appear because we used the
spherical coordinates to describe the cloud. At very large radii,
where the sphericalness is negligible (i.e. $r\rightarrow \infty$),
these coefficients reduce to the coefficients of the characteristic
equation of the well-known pioneered work of [1]. Using
non-dimensional quantities
\begin{equation}\label{nonq}
    y \equiv \frac{\omega}{kc}, \quad \sigma_T \equiv
    \frac{\nu_T}{kc}, \quad \sigma_\rho \equiv \frac{\nu_\rho}{kc},
    \quad \lambda \equiv \frac{1}{kr},
\end{equation}
the characteristic equation can be written as
\begin{equation}\label{chsis}
    y^3 + \sigma_T y^2 + \left(1 - 2 i \lambda -
    \frac{2\lambda^2}{\gamma} \right) y + \left[ \frac{1}{\gamma} \left( \sigma_T - \sigma_\rho\right)
    - i \frac{2\lambda}{\gamma} \sigma_T - \frac{2\lambda^2}{\gamma} \sigma_T \right] = 0.
\end{equation}
Here, we use the Laguerre method (e.g., [16]) to find numerically the roots of the characteristic equation (\ref{chsis}). The results for instability in the $\sigma_T$-$\sigma_\rho$ plane are shown in Fig.~\ref{sisfig}. Note that $\lambda=0$ corresponds to the results of [1], with the usual isobaric and isentropic instability criteria as $\nu_\rho > \nu_T$, and $\nu_\rho < -(\gamma-1) \nu_T$, respectively. The results show that increasing of $\lambda$ (i.e., perturbations with longer wavelengths at each radius $r$) decreases the stability region in the $\sigma_T$-$\sigma_\rho$ plane. The importance of linear TI can be expressed by the growth rate, $\Re(\omega)$, of unstable regions through the $\sigma_T$-$\sigma_\rho$ plane. Finding the roots of (\ref{chsis}) for different values of $\lambda$ shows that $\Re(y/\lambda)$ is a decreasing function of $\lambda$. Thus, at each radial position $r$, perturbations with shorter wavelengths (i.e., smaller $\lambda$) have greater growth rates, which corresponds to more chance for thermally unstable growth.

\section{Summary and conclusions}

TI is an important trigger mechanism that can explain the formation of density condensations through some regions of interstellar clouds and circumgalactic medium. In this paper, we turned our attention to a simple model as a spherical cloud and investigated the occurrence of linear TI through it. First, we surveyed thermally equilibrium models of the quasi-static spherically symmetric clouds; the radial profiles of density, temperature, pressure, and enclosed mass are shown in Fig.\ref{prof}. After that, we used the perturbation method to investigate the occurrence and growth rates of linear TI through these equilibrium models.

To represent the thermal balance case, a parametric relation between density and temperature as $T = \rho^\eta$ is used. Fig.~\ref{prof} shows that in all models, density decreases with increasing radius, but its slope depends on the parameter $\eta$. The same behavior can be seen in the pressure profile. For the models that temperature increases with density increase ($\eta>0$), the pressure drop versus radius is more than the models that temperature decreases with density increase ($\eta<0$).

To investigate TI, we considered the simplest case in which $\eta=0$. In this case, the characteristic equation (\ref{charac}) reduces to a simple form (\ref{chsis}), which by ignorance of the sphericalness (i.e., $r\rightarrow \infty$ so that $\lambda\rightarrow 0$), will be reduced to the characteristic equation of the well-known work of Field~(1965). The
Fig~\ref{sisfig} shows the thermally unstable regions in the$\sigma_T$-$\sigma_\rho$ plane; considering the sphericalness of the cores (i.e., greater values of $\lambda$) results in increasing the instability regions in this plane. Also, comparing the real parts of the roots of (\ref{chsis}) for different values of $\lambda$ demonstrates that at each radius $r$, perturbations with smaller wavelengths become more thermally unstable (i.e., have greater growth rates) than longer ones.
\\[5mm]

\noindent \Large\textbf{References}\\[5mm]
\small \noindent
1.~~Field, G. B. (1965). Thermal Instability. ApJ,
142, 531. DOI: 10.1086/148317
\\[3mm]
\noindent
2.~~Hunter, J. H. (1966). The role of thermal
instabilities in star formation. MNRAS, 133, 239. DOI:
10.1093/mnras/133.2.239
\\[3mm]
\noindent
3.~~Fukue, T. \& Kamaya, H. (2007). Small Structures via
Thermal Instability of Partially Ionized Plasma: I. Condensation
Mode. ApJ, 669, 363. DOI: 10.1086/521268
\\[3mm]
\noindent
4.~~Nejad-Asghar, M. (2011). Formation of low-mass
condensations in molecular cloud cores via thermal instability.
MNRAS, 414, 470. DOI: 10.1111/j.1365-2966.2011.18412.x
\\[3mm]
\noindent
5.~~Choudhury, P. P. \& Sharma, P. (2016). Cold gas in
cluster cores: global stability analysis and non-linear simulations
of thermal instability. MNRAS, 457, 2554. DOI: 10.1093/mnras/stw152
\\[3mm]
\noindent
6.~~Nejad-Asghar, M. (2019). Thermal instability through
the outer half of quasi-static spherically symmetric molecular
clumps and cores. Ap\&SS, 364, 122. DOI: 10.1007/s10509-019-3616-y
\\[3mm]
\noindent
7.~~Dannen, R. C., Proga, D., Waters, T. \& Dyda, S.
(2020). Clumpy AGN Outflows due to Thermal Instability. ApJ, 893,
34. DOI:10.3847/2041-8213/ab87a5
\\[3mm]
\noindent
8.~~Antolin, P., Mart\'{\i}nez-Sykora, J. \& \c{S}ahin, S,
(2022). Thermal Instability-Induced Fundamental Magnetic Field
Strands in the Solar Corona. ApJ, 926, 29. DOI:
10.3847/2041-8213/ac51dd
\\[3mm]
\noindent
9.~~Gronkiewicz, D., R\'{o}\.{z}a\'{n}ska, A., Petrucci, P.
\& Belmont, R, (2023). Thermal instability as a constraint for warm
X-ray coronas in active galactic nuclei. A\&A, 675,198. DOI:
10.1051/0004-6361/202244410
\\[3mm]
\noindent
10.~~Hermans, J. \& Keppens, R. (2024). A spectroscopic
investigation of thermal instability for cylindrical equilibria with
background flow. A\&A, 686, 180. DOI: 10.1051/0004-6361/202348337
\\[3mm]
\noindent
11.~~Neufeld, D. A., Lepp, S. \& Melnick, G. J. (1995).
Thermal Balance in Dense Molecular Clouds: Radiative Cooling Rates
and Emission-Line Luminosities. ApJS, 100, 132. DOI: 10.1086/192211
\\[3mm]
\noindent
12.~~Glassgold, A. E. \& Langer, W. D. (1973). Cosmic-Ray
Heating and Molecular Cooling of Dense Clouds. ApJ, 179, 147. DOI:
10.1086/181137
\\[3mm]
\noindent
13.~~Li, P., Myers, A. \& McKee, C. F. (2012). Ambipolar
Diffusion Heating in Turbulent Systems. ApJ, 760, 33. DOI:
10.1088/0004-637X/760/1/33
\\[3mm]
\noindent
14.~~Wiener, J., Zweibel, E. G. \& Ruszkowski, M. (2019).
Cosmic ray acceleration of cool clouds in the circumgalactic medium.
MNRAS, 489, 205. DOI: 10.1093/mnras/stz2007
\\[3mm]
\noindent
15.~~Wiersma, R. P. C., Schaye, J.\& Smith, B. (2009). The
effect of photoionization on the cooling rates of enriched,
astrophysical plasmas. MNRAS,393, 99. DOI:
10.1111/j.1365-2966.2008.14191.x
\\[3mm]
\noindent 16.~~Press, W. H., Teukolsky, S. A., Vetterling, W. T. \&
Flannery, B. P. (1992). Numerical recipes in FORTRAN: The art of
scientific computing (2th ed.). Cambridge University Press.
\\[3mm]
\scriptsize\-----------------------------------------------------------------------------------------------------------------------------------------\\\copyright \it 2024  Author; This is an Open Access article distributed under the terms of the Creative Commons Attribution License http://creativecommons.org/licenses/by/2.0,  which permits unrestricted use, distribution, and reproduction in any medium,
provided the original work is properly cited.

\end{document}